\newcommand\dns{D$_{\rm n} - \sigma$}
\newcommand\kms{$\rm\,km\,s^{-1}\,$}
\newcommand\kmsm{$\rm\,km\,s^{-1}Mpc^{-1}\,$}
\newcommand\etal{{\it et al.}}
\newcommand\re{$r_e$}
\newcommand\mue{$\mu_e$}

\documentstyle[11pt,aaspp4]{article}

\slugcomment{(to appear in the Astronomical Journal)}

\lefthead{Scodeggio, Giovanelli \& Haynes}
\righthead{Photometric Fundamental Plane}

\begin{document}

\title{An Economical Technique for the Estimate of Galaxy Distances:
the Photometric Fundamental Plane}

\author{Marco Scodeggio\altaffilmark{1}, Riccardo Giovanelli and Martha P. 
Haynes}
\affil{Center for Radiophysics and Space Research and National Astronomy
and Ionosphere Center\altaffilmark{2}, Cornell University, Ithaca, NY 14853\\
mscodegg@eso.org \\ riccardo, haynes@astrosun.tn.cornell.edu}

\altaffiltext{1}{Present address: European Southern Observatory, 
Karl-Schwarzschild-Stra$\beta$e 2, D-85748, Garching bei M{\"u}nchen, 
Germany.}
\altaffiltext{2}{The National Astronomy and Ionosphere Center is
operated by Cornell University under a cooperative agreement with the
National Science Foundation.}

\begin{abstract}
We show that it is possible to define a purely Photometric Fundamental 
Plane (PFP) for early--type galaxies. This relation is similar to the standard 
Fundamental Plane (FP), and is obtained by replacing the velocity dispersion 
parameter with the difference between the magnitude of a galaxy and that of 
the mode of the Gaussian luminosity function of E and S0 galaxies.
The use of magnitude differences as a third parameter allows a significant 
reduction in the dispersion of the PFP relation when compared to the Kormendy 
relation between effective radius and effective surface brightness, but limits 
the application of this method to galaxies in clusters.
The dispersion of $\sim 0.10$ in $\log R_e$ about the mean plane in the
PFP is comparable to that of the standard FP. However, the use of the mode of 
the luminosity function to compute the magnitude differences introduces a 
systematic uncertainty in the derivation of the PFP relation zero-point, so 
that its accuracy for distance determinations does not scale with the square 
root of the number of objects used to perform the fit. The method is also 
vulnerable to any bias that might affect the estimate of the mode of the 
luminosity function.
If however the mode of the luminosity function can be reliably determined,
the PFP relation can provide distance estimates with an accuracy comparable to 
the FP relation, with the advantage that the use of photometric parameters 
alone reduces drastically the observational requirements of the PFP, in 
comparison with those of the FP relation. This practical advantage makes the 
PFP a very economical distance indication method.
\end{abstract}

\section{Introduction}

Techniques for the measurement of redshift--independent distances play a
fundamental role in cosmology. Ideally, they would rely on the use of a 
``standard candle'', i.e. a class of easily discernible objects of constant 
intrinsic properties. In most extensively applied techniques, e.g. those
that use the Tully--Fisher (TF) relation for spirals (Tully \& Fisher 1977) 
and the \dns\ or Fundamental Plane (FP) relation for spheroidals (Dressler 
\etal\ 1987; Djorgovski \& Davis 1987), the standard candle concept is 
replaced by an empirically calibrated scaling relation (see also Jacoby 
\etal\ 1992 for a review). TF, \dns\ and FP all correlate a photometric, 
distance-dependent parameter with a kinematic, distance-independent one.
Distance\footnote{Because of the uncertainty on the value of the 
Hubble constant, and because the number of nearby galaxies that could be used 
for an absolute calibration of the template relations is very small, the TF and 
FP relations are most commonly used to derive relative distances. 
Therefore here the term ``distance'' is used as an abbreviation for the more 
appropriate ``recession velocity corrected for peculiar motions''.} estimates 
for a single galaxy are obtained with a typical uncertainty that varies between 
12 and 25\%. Because of this relatively high accuracy, these relations
have been extensively used in mapping deviations from the universal expansion 
up to $cz \sim 10,000$ \kms. More recently they also have been used to study the 
evolution of the stellar population in high-redshift galaxies ($z$ from 0.2 to 0.6), 
taking advantage of the dependence of the zero-point of both the TF and the FP 
relation on the galaxy mean mass-to-light ratio (see for example Vogt \etal\ 1993, 
Bershady 1996, van Dokkum \& Franx 1996, Vogt \etal\ 1996).

From the observational view-point, the ingredients of TF, \dns\ and FP are
analogous. Magnitudes (for TF), radii and surface brightnesses (for \dns\ 
and FP) require optical photometric data of similar quality. For a given nearby
target ($z<0.1$), a ten minute exposure of a high efficiency CCD on a 1m class 
telescope generally suffices. Rotational widths (for TF) or velocity 
dispersion (for \dns\ and FP) are, on the other hand, much more demanding. 
Larger telescopes are required than those necessary for the photometry, 
as well as significantly longer exposures. While the photometry of a TF, 
\dns\ or FP program can be economically acquired, the spectroscopy requires 
an observational investment of roughly one order of magnitude higher cost.

An economical and ubiquitously applicable distance-indication method, 
based entirely on photometric parameters, would be an extremely valuable tool
for observational cosmology, if it were to achieve an accuracy comparable to
those characteristic of the TF and FP relations.
The Kormendy relation (Kormendy 1977) has been used as an economical
substitute for the FP relation (Pahre \etal\ 1996). However its large scatter, 
and the fact that the residuals in the relation correlate with the galaxy 
luminosity (or, equivalently, with the galaxy velocity dispersion), make this 
relation extremely sensitive to bias effects, such as the cluster population 
incompleteness bias for cluster samples, or the Malmquist bias for field 
samples.
Other extensively used photometric distance indicators are based on the 
determination of a characteristic magnitude, like the magnitude of the brightest 
galaxy in a cluster (Sandage 1972, Sandage \& Hardy 1973, Hoessel 1980, 
Lauer \& Postman 1994), or the peak of a Gaussian luminosity function (LF). 
The latter has been commonly employed using globular clusters or planetary nebulae 
within a single galaxy (see Jacoby \etal\ 1992), but it could be used with cluster 
E and S0 or spiral galaxies, that have a Gaussian LF (Sandage \etal\ 1985). 
Similarly, the LF of the global galaxy population in a cluster could be used, 
adopting the magnitude M$_*$ in a Schechter LF (Schechter 1976) as a standard 
candle. The reliability of this application is obviously dependent on the degree 
of universality of the LF for clusters of galaxies.
Still unexplored is another purely photometric, and potentially very useful,
technique, based on a family of relations between a galaxy effective radius, 
effective surface brightness, and broadband colors discovered by de Carvalho 
\& Djorgovski (1989).

Here we propose to use jointly the Kormendy relation and any one of the methods 
that can provide a characteristic magnitude for a cluster sample, to derive a 
modified version of the FP relation, based entirely on photometric parameters.
We define for each galaxy $\Delta M$ as the difference between the galaxy's 
magnitude and the sample characteristic magnitude, and use $\Delta M$, a 
distance--independent parameter, in substitution of the velocity dispersion in 
the FP relation. We illustrate the characteristics of such a Photometric 
Fundamental Plane (PFP) with a sample of 405 early-type galaxies in the Coma, 
A1367, and A2634 clusters. The characteristic magnitude for each cluster sample 
is obtained from the peak of the Gaussian LF that best approximates the sample 
magnitude distribution. 
This procedure is quite similar to the one previously used by de Carvalho 
\& Djorgovski (1989), with the difference that those authors used broadband colors
as a substitute for the velocity dispersion in the FP relation.
The main advantage of the PFP relation is the improved accuracy it provides 
with respect to both the Kormendy relation and the method used to derive the 
sample characteristic magnitude. Such accuracy however depends on the 
reliability of the characteristic magnitude.
First, and most important, any observational bias that affects the measurement 
of this magnitude will affect the PFP relation as well. Second, the uncertainty 
with which this magnitude is derived becomes a systematic uncertainty in the 
derivation of the PFP zero point. Therefore the accuracy of this derivation 
does not scale with the square root of the number of objects used in the fit, 
but is limited to a fraction of the uncertainty in the determination of the 
characteristic magnitude.

This paper is organized as follows: in section 2 we discuss the data used to 
obtain the PFP relation. The derivation of the PFP and its accuracy for
distance determinations are discussed in section 3, 
while the discussion and conclusions are in section 4. Distance-dependent 
parameters are computed assuming H$_\circ$=100~h~\kmsm.

\section{The data}

As part of a study aimed at the cross--calibration of the TF and FP techniques, 
we obtained I band CCD images of early-type galaxies in 10 nearby clusters, 
including Coma, A1367, and A2634, with the 0.9m telescope of the Kitt Peak 
National Observatory (KPNO), during 3 observing runs between 1994 April, and 
1995 September. The telescope was used with the f/7.5 secondary, field
corrector and T2KA CCD chip (2048 x 2048 pixels), to obtain a field of view 
of 23\arcmin\ x 23\arcmin, with a spatial scale of 0.68\arcsec\ per pixel. 
The median effective seeing (the median FWHM of the stellar profiles) for these
observations was 1.6\arcsec. All frames were obtained with 600 seconds 
integration time. Observations of Landolt fields (Landolt 1992), both at I and 
at R band, were repeated many times during each night, at airmasses between 
1.2 and 2.5, to obtain the photometric zero point calibration and the 
atmospheric extinction coefficient. The mean uncertainty in the zero point 
calibration at I band was 0.021 magnitudes. 

Complete details on the observations and the data reduction procedure will be 
presented elsewhere (Scodeggio 1997, Scodeggio \etal\ 1997b). Here we briefly 
summarize those details. All CCD frames were reduced using standard 
IRAF\footnote{IRAF (Image Reduction and Analysis Facility) is distributed by 
NOAO, which is operated by the Association of Universities for Research in 
Astronomy, Inc. (AURA), under cooperative agreement with the National Science 
Foundation.} procedures, and surface photometry measurement were obtained using 
the GALPHOT surface photometry package written for IRAF/STSDAS\footnote{STSDAS
(Space Telescope Science Data Analysis System) is distributed by the Space 
Telescope Science Institute, which is operated by AURA, under contract to the 
National Aeronautics and Space Administration.} by W. Freudling, J. Salzer, and 
M.P. Haynes. All frames were bias-subtracted, flat-fielded, and sky 
background-subtracted using the mean number of counts measured in 10-12 regions 
of ``empty'' sky. The uncertainty in the sky background is typically 0.2\%. All 
pixels contaminated by the light of foreground stars or nearby galaxies, or by 
cosmic rays hits, were blanked, and excluded from the final steps of surface 
photometry.

Photometric measurements were obtained for all early-type galaxies with an 
available redshift measurement, and that are to be considered cluster members 
according to the criteria described by Giovanelli \etal\ (1997a). In addition 
a small number (64 galaxies out of the total 405) of galaxies that do not have 
redshift measurements available was included in the sample, because their size 
and luminosity make them likely cluster members.
The 2-dimensional light distribution of each galaxy was fitted with elliptical 
isophotes, using a modified version of the STSDAS {\it isophote} package,
maintaining as free parameters the ellipse center, ellipticity and position 
angle, and incrementing the ellipse semi-major axis by a fixed fraction of 
its value at each step of the fitting procedure. The fitted parameters  
yield a model of the galaxy light distribution, which is used to compute 
integrated magnitudes as a function of semi-major axis.
For each galaxy the effective radius $r_e$ and the effective surface brightness 
$\mu_e$ (the mean surface brightness within $r_e$) were obtained by fitting the 
radial surface brightness profile with a de Vaucouleurs $r^{1/4}$ law.
The fit was performed from a radius equal to twice the seeing radius, out to 
the outermost isophotes for E galaxies;  for S0 and S0a galaxies only the 
central core was fitted. The median uncertainty on the determination of $r_e$ 
and $\mu_e$ is 5\% and 0.06 mag., respectively. Total magnitudes were obtained 
independently from the $r^{1/4}$ fit, by extrapolating the $r^{1/4}$ fit to 
infinity (E galaxies), or by extrapolating to infinity the exponential profile 
that fitted the outer parts of the galaxy light profile (S0 and S0a galaxies), 
and adding the flux corresponding to the extrapolated part of the profile to 
the one measured within the outermost fitted galaxy isophote. The median 
uncertainty in the determination of the total magnitude is 0.06 mag.

Standard corrections were applied for Galactic extinction, using Burstein \& 
Heiles (1978)  prescriptions, and $A_I = 0.45A_B$, for the cosmological 
k-correction term (which was taken to be $2.5\log(1+z)$ because of the flat 
spectrum of early-type galaxies in the far red), and for the surface brightness 
$(1+z)^4$ cosmological dimming. Both \re\ and \mue\ were corrected for the 
effects of seeing following the prescriptions of Saglia \etal\ (1993, see in 
particular their figure 8).

\section{The Photometric Fundamental Plane}

\subsection{Defining the relation}

Well known correlations exist between early--type galaxy properties, such as 
that between luminosity and: \re\ (Fish 1964), $\sigma$ (Faber \& Jackson 1976),
and \mue\ (Binggeli \etal\ 1984), and the relation between \re\ and 
\mue\ (Kormendy 1977). These correlations exhibit much larger scatter than can 
be accounted by measurement errors alone. The idea that early-type galaxies 
populate a plane in the 3--parameter space (\re, \mue, $\sigma$), independently 
introduced by Djorgovski \& Davis (1987) and by Dressler \etal\ (1987), led to 
a significantly reduced scatter with respect to the Faber--Jackson or Kormendy 
relation, and made possible its use as a redshift-independent distance 
indicator.

Clusters of galaxies have been favorite targets for FP studies because they 
provide both an environment rich in early-type galaxies and large samples of 
objects all roughly at the same distance. Recent studies of that nature include 
those of Guzm{\'a}n \etal\ (1993), J{\o}rgensen \etal\ (1996) 
and our own (Scodeggio \etal\ 1997a). There is remarkable quantitative 
agreement in the FP calibration of those studies. In particular, the last two 
show that the r.m.s. scatter of the residuals in $\log r_e$, is $\simeq$0.085, 
or 20\% uncertainty on the distance. For comparison, the Tully-Fisher relation 
has a dispersion of $\simeq$0.25--0.45 magnitudes (equivalent to a distance 
uncertainty of 12--24\%, depending on the galaxy's rotational velocity; 
Giovanelli \etal\ 1997b).

The Kormendy (1977) relation between \re\ and \mue\ has been recently 
used by Pahre \etal\ (1996) as an economical substitute of the FP in two high
redshift clusters, to obtain an improved version of the classical Tolman test 
for the expansion of the Universe. The scatter in the Kormendy relation is 
fairly large, and little use can be found for it as a distance indicator for 
single galaxies. However the disadvantage of the large scatter can be partly 
offset by the advantage of having a relation between two purely photometric 
parameters, when the relation is applied to clusters of galaxies. In a rich 
cluster, the large number of objects can statistically 
compensate for the large scatter in the relation in deriving the
cluster distance. In Figure 1 we present the Kormendy relation for 405 E and S0 
galaxies in the Coma, A1367, and A2634 clusters. The least squares, direct 
linear fit to this data set\footnote{This is likely to be a biased estimate of the 
true relation between $\log R_e$ and $\mu_e$, because the fitting does not take 
into proper consideration the correlation between the measurement errors for the 
two parameters, but a detailed discussion of the Kormendy relation is beyond the 
purpose of this paper.} is given by 
\begin{equation}\label{Korm}
\log R_e = 0.284 (\mu_e - 19.45) + 0.495 - \log h
\end{equation}
where $R_e$ is in kiloparsec, and is computed placing each galaxy at the 
distance indicated by the cluster redshift, in the CMB reference frame. In 
doing this we are ignoring the possible effect of peculiar motions on the 
derivation of $R_e$, because
the peculiar velocities of Coma, A1367 and A2634 have been shown to be quite
small (Giovanelli \etal\ 1997b, Scodeggio \etal\ 1997a). The r.m.s. scatter in 
$\log R_e$ is 0.19, equivalent to an uncertainty of 0.95 mag in the distance 
modulus of a single galaxy. However, the statistical uncertainty in the fit 
zero point, because of the large sample being used, is only 0.010, equivalent 
to 0.05 mag, or to a distance uncertainty of 170 \kms\ at the distance to the 
Coma cluster.

Figure 2 shows the well known fact that the residuals from the Kormendy 
relation (\ref{Korm}) are not random: they correlate very well with the galaxy 
magnitude, or, equivalently, with the galaxy velocity dispersion, bright 
galaxies exhibiting systematically positive residuals. We remark that the 
magnitudes used here are derived directly from the observed galaxy light 
distributions, and are therefore measured independently from the \re\ and \mue\ 
parameters. Because of the combination of the large scatter with residuals that 
are correlated with the galaxy luminosity, the Kormendy relation is severely 
affected by the cluster population incompleteness bias (Teerikorpi 1987, 
Sandage 1994a,b). The measured zero point and dispersion of the
Kormendy relation depend on the limiting magnitude of the sample, as can be 
inferred from Figure 2 if we imagine the removal of all galaxies fainter than a 
certain limit. It is therefore very important to derive accurate bias 
corrections before using the Kormendy relation for cosmological applications 
like redshift-independent distance measurements.

It is clear that the two correlations shown in Figure 1 and 2 can be combined to
produce a global relation in the 3-parameter space of $\log R_e$, $\mu_e$, and 
$M$. This could be used, in principle, as a distance indication relation, where 
$\mu_e$ would be the distance-independent parameter, and $\log R_e$ and $M$ 
would be the distance-sensitive parameters, derived from the observed 
$\log r_e$ and $m$ assuming all galaxies are at the distance indicated by the 
cluster redshift. Unfortunately, the best fitting plane to the $\log R_e$, 
$\mu_e$, $M$ relation is given by 
\begin{equation}\label{mueremag}
\log R_e = -0.13~M + 0.26~\mu_e + const.
\end{equation}
and the combination of distance-dependent parameters $\log R_e + 0.13~M$ is 
quite close to the distance-independent combination 
$\log R_e +0.2~M = \log r_e + 0.2~m$.
Equivalently, the best fitting plane (\ref{mueremag}) is almost parallel to the 
distance-independent plane
\begin{equation}\label{deva}
m = \mu_e + 5 \log r_e + const.
\end{equation}
that would be populated by a perfectly homologous family of early-type galaxies,
making the relation (\ref{mueremag}) of little practical use for distance 
determinations. However, if a characteristic magnitude could be defined 
in a distance-independent way for each cluster sample, this relation
could be re-written in terms of the magnitude difference $\Delta M$ with 
respect to the characteristic magnitude, instead of the galaxy total magnitude. 
In this way a relation completely analogous to the FP relation could be 
obtained, using only photometrically derived parameters, since magnitude 
differences are distance-independent quantities.

\subsection{The characteristic magnitude}

A number of methods are available for the determination of a characteristic 
magnitude for a cluster sample, using only photometric data. Sandage (1972), 
and Sandage \& Hardy (1973) have used the luminosity of the first ranked galaxy 
in a cluster to study the linearity of the Hubble flow out to very large 
distances. These authors claim that all such galaxies have approximately the 
same luminosity, with a dispersion of only 0.28--0.32 magnitudes (depending on 
the sample used). A modified version of this distance-estimation method was 
proposed by Hoessel (1980), and has been used most recently by Lauer \& Postman 
(1994) to study the motion of Abell clusters with $cz \leq 15,000$ \kms.
The uncertainty associated with this method has been estimated to be 
approximately 16\%, or 0.35 magnitudes (Lauer \& Postman 1994).
Also LF's have been used as distance indicators. Well known is
the use of the globular clusters and planetary nebulae LF (see Jacoby
\etal\ 1992, and references therein) to derive redshift-independent distances 
for individual galaxies. The same technique can be applied to the galaxy LF
in a cluster, provided that such LF is universal. 
Since giant E and S0 galaxies and spiral galaxies have been shown to
have a Gaussian LF (Sandage \etal\ 1985), its peak $M_{peak}$ can provide an 
accurate estimate of a characteristic magnitude. Similarly, fitting a Schechter 
LF (Schechter 1976) to the entire cluster population might provide a 
characteristic magnitude, in terms of the parameter $M_*$. Typical statistical
uncertainties associated with the determination of $M_{peak}$ and $M_*$ are 
0.2 and 0.3 magnitudes, respectively (see, for example, Jacoby \etal\ 1992, 
and Marzke \etal\ 1994).

Here we use a Gaussian fit to the LF of E and S0 galaxies to
determine the characteristic magnitude for our cluster samples.
The fit is performed using a maximum likelihood method (Malumuth \& Kriss 1986),
combining the 3 cluster samples, and excluding the brightest galaxy in each 
cluster from the fit. As in the case of the Kormendy relation, we compute 
absolute magnitudes assuming that each galaxy is at the distance indicated by 
the cluster redshift, in the CMB reference frame, and ignoring the effect of 
possible peculiar motions (known, as we said, to be quite small) on the 
redshift-distance conversion. The best fitting Gaussian LF yields 
$M_{peak} = -20.55$, with a dispersion of 1.25 magnitudes. This is in good 
agreement with the location of  the peak observed in the Coma cluster LF by 
Biviano et al. (1995) and with the Gaussian LF obtained by Sandage \etal\ (1985) 
for a complete sample of E and S0 galaxies in the Virgo cluster, when an average 
(B--I) color of  $\simeq 2.15$ for early-type galaxies is assumed. 
Figure 3 shows the results of the maximum likelihood fitting. In Fig. 3a the 
stair-step line represents the observed integral magnitude distribution, while the 
smooth curve is derived integrating the best fitting Gaussian. The inset show the 
best fitting Gaussian parameters, and the 68\% (1$\sigma$) and 95\% (2$\sigma$) 
joint confidence contours for those parameters. In Fig. 3b the histogram shows the 
differential magnitude distribution, with bins of width 0.4 mag, and the solid line 
curve shows the best fitting Gaussian.
The completeness limit for the sample is shown by the vertical dashed line.
The statistical uncertainty in the determination of the Gaussian peak is
$\simeq 0.22$ mag. However it still remains to be demonstrated that the location
of this peak does not change systematically as a function of cluster parameters
like the dynamical evolution state, the richness, or the density of the
intra-cluster medium.

\subsection{The PFP and its accuracy}

Defining $\Delta M = M - M_{peak}$, we see the E and S0 galaxies populating a 
plane in the purely photometric 3-parameter space  ($\log R_e$, \mue, 
$\Delta M$), which we term the Photometric Fundamental Plane or PFP. 
Figure 4 shows an edge-on view of the 
PFP for the combined sample of Coma, A1367, and A2634 E and S0 galaxies.
The best fitting plane, obtained averaging the results of the 3 possible fits 
that can be performed using one parameter as the dependent variable and the 
remaining two as the independent ones, is given by the relation
\begin{equation}\label{PFP}
\log R_e = -0.13 ~\Delta m + 0.264 (\mu_e - 19.45) + 0.464 - \log h
\end{equation}
We remark that the coefficients derived here for equation (\ref{PFP}) should be 
considered only provisional, because of the difficult statistical problem of 
fitting a relation among parameters that do not have zero covariance. From this 
point of view the PFP case is very similar to the FP one, because measurement 
errors in $\log R_e$ and \mue\ are strongly correlated, but they have negligible 
correlation with the measurement errors in $\Delta M$, because total magnitudes 
are determined independently from the $r^{1/4}$ fit used to determine $\log R_e$ 
and \mue.

A hint of non-linearity in the relation is apparent. However, the use
of a quadratic relation produces a fit that is not significantly better than the
one obtained using equation (\ref{PFP}). A much larger sample will be
required to settle this point.
The r.m.s. scatter in $\log R_e$ about the PFP plane is 0.096, which is 
approximately half the scatter shown by the Kormendy relation, and very similar 
to the scatter shown by the FP relation, as discussed above.
This scatter, however, is not the only source of uncertainty in the 
determination of a cluster distance with the PFP. The uncertainty with which 
the characteristic magnitude is determined introduces a systematic uncertainty 
in the determination of the PFP zero-point, that must be added to the 
statistical uncertainty produced by the scatter. Because this last term scales
approximately with the square root of the number of data-points, for a large 
sample the systematic uncertainty due to the characteristic magnitude 
determination becomes the dominant source of error in a cluster distance 
determination.
However, the value of the $\Delta M$ coefficient in the PFP is such that the 
effect of this systematic uncertainty on distance determinations performed with 
the PFP is less severe than it would be on the equivalent determination based 
solely on the characteristic magnitude. In the latter case the distance 
uncertainty $\sigma_d$ is related to the characteristic magnitude uncertainty 
$\sigma_M$ by the usual $\sigma_d / d = 0.2\ln 10~\sigma_M$, whereas in the PFP 
case the relation (considering only the characteristic magnitude contribution) 
is $\sigma_d / d = 0.13\ln 10~\sigma_M$, which is $\simeq 1.5$ times smaller.

Figure 5 shows a comparison of the accuracy in distance determinations that can
be obtained using the Fundamental Plane, the PFP, and the Kormendy relation, as
a function of sample size. We assume that in all cases statistical 
uncertainties scale with the square root of the number of objects used in the 
fit. The horizontal dashed lines give the accuracy achieved
using the FP with a fixed sample size of 10, 20, or 30 galaxies, typical of the 
samples currently being used for distant and nearby clusters. The scatter in 
the FP is assumed to be 0.085 (0.43 mag in distance modulus; see J{\o}rgensen 
\etal\ 1996, and Scodeggio \etal\ 1997a). The two solid line curves give the 
accuracy achieved using the PFP, for two different values of the uncertainty 
associated with the determination of the characteristic magnitude (0.25 mag for 
the upper curve, and 0.15 mag for the lower curve), for a scatter in the PFP of 
0.096 (0.48 mag in distance modulus). The dotted line gives the accuracy 
obtained with the Kormendy relation, for a scatter of 0.19 (0.95 mag in 
distance modulus), and assuming that the sample completeness bias 
corrections do not introduce further uncertainties. In this sense, and also
because our sample is incomplete at the fainter magnitudes and therefore
we are under-estimating the true scatter in the Kormendy relation, this line 
should be considered an over-optimistic estimate of the accuracy achievable 
with the Kormendy relation.

\section{Discussion and Conclusions}
As in the case of the relations described by de Carvalho \& Djorgovski (1989),
the PFP can be understood as a generalization of the fundamental plane relation
of early-type galaxies. Although our understanding of the processes of galaxy 
formation and evolution is still relatively limited, the mass of a galaxy 
appears to be the fundamental regulating parameter. 
Because of this, many observable quantities, like the galaxy 
luminosity, velocity dispersion, metallicity, and broad-band colors, are 
correlated with the mass, and therefore with each other. 
The precise form of those correlations is not always well constrained, 
but a subset exists, including the FP, the PFP and the
relation suggested by de Carvalho \& Djorgovski,
that appears to yield a tighter description of 
early-type galaxies, within the limits imposed by measurement uncertainties 
and the intrinsic scatter present among FP-like relations.

The PFP can be a useful tool for observational cosmology. Its most positive
characteristics are the relatively modest observational requirements, combined 
with its potential accuracy. The PFP relation requires only the photometric 
observations necessary also for the FP relation, without any need for 
spectroscopic observations. 
The characteristic magnitude required by the method can be derived from the 
photometric observations as well, in the form of the mode of the LF, or 
directly using the magnitude of the n$^{\rm th}$ brightest cluster member. 
This requirement, however, limits the applicability of the method to clusters 
of galaxies, and places the ultimate limit to its accuracy.

The accuracy of the PFP zero-point is constrained by the uncertainty in the 
measurement of the characteristic magnitude. We have seen that, at best, the 
uncertainty in the PFP zero point can be $\sim$60\% of the uncertainty in the 
characteristic magnitude. Moreover, the PFP is affected not only by the bias 
intrinsic to the derivation of a template PFP relation, but its zero point is 
sensitive to biases that might affect the characteristic magnitude. 
Correlations are known to exist between the luminosity of the brightest galaxy 
in a cluster and the cluster Bautz-Morgan type (Sandage \& Hardy 1973), and 
also the cluster X-ray luminosity (Hudson \& Ebeling 1997). The universality 
of the LF for clusters is still very much debated. 
Therefore extreme care will be required for the application of the PFP
for distance determination. However the properties of the method used to derive
the characteristic magnitude could be in principle derived from the same data
used to obtain the PFP relation, and should provide the opportunity to perform
stringent consistency checks.

If the characteristic magnitude can be constrained to within 0.3 magnitudes or 
less, the PFP can offer an accuracy in distance measurements comparable to that 
of the FP, for a typical FP cluster sample, requiring only one fifth to one 
tenth of the telescope time. Also important is the fact that the imaging 
observations do not require a 4-5 meter class telescope (or 8-10 meter 
telescope, for high redshift clusters), but can be obtained with a 1 meter 
class telescope.

Finally, one important point must be made regarding the use of the PFP relation
at low and high redshift. At low redshift, where the relation would be used to
measure deviations from the Hubble flow, the characteristic magnitude must be
derived separately for each cluster. At high redshift, instead, where the 
relation might be used to study evolutionary changes in the mass-to-light ratio 
of early-type galaxies, peculiar velocities have a negligible effect on the 
characteristic magnitude. Different cluster samples at the same redshift could 
thus be combined, improving the determination of the characteristic magnitude, 
and the accuracy of the PFP method.

\acknowledgments

We would like to thank George Djorgovski, the referee, for his valuable and
constructive criticism, that helped improve this paper. MS has benefitted
from useful discussions with Enzo Branchini, Bianca Garilli, Guido
Chincarini, and Ralf Bender.
The results presented in this paper are based on observations carried out at
the Kitt Peak National Observatory (KPNO). KPNO is operated by Associated
Universities for Research in Astronomy, under a cooperative agreement with 
the National Science Foundation.  This research is part of the Ph.D. Thesis 
of MS, and is supported by the NSF grants AST94--20505 to RG and AST92--18038 
to MPH.


\newpage

\newpage
\pagestyle{empty}
\voffset=2truecm
\
\plotone{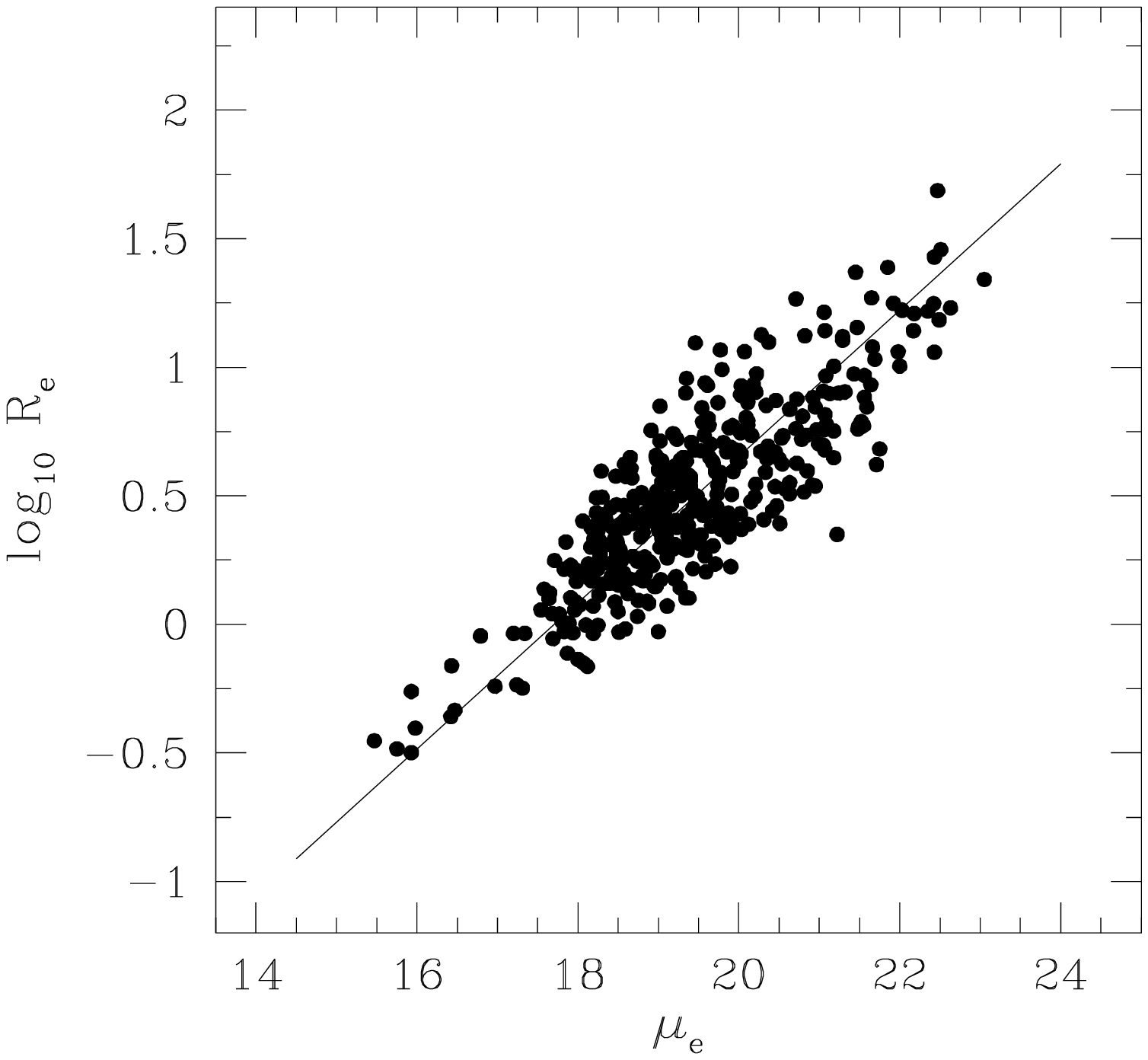}
\figcaption[fig1.ps]{The Kormendy relation between the logarithm of the 
effective radius (in kiloparsec) and the effective surface brightness, 
for 405 early-type galaxies in the Coma, A1367, and A2634 clusters. The solid 
line is the best fit to the correlation (equation 1).}

\newpage

\plotone{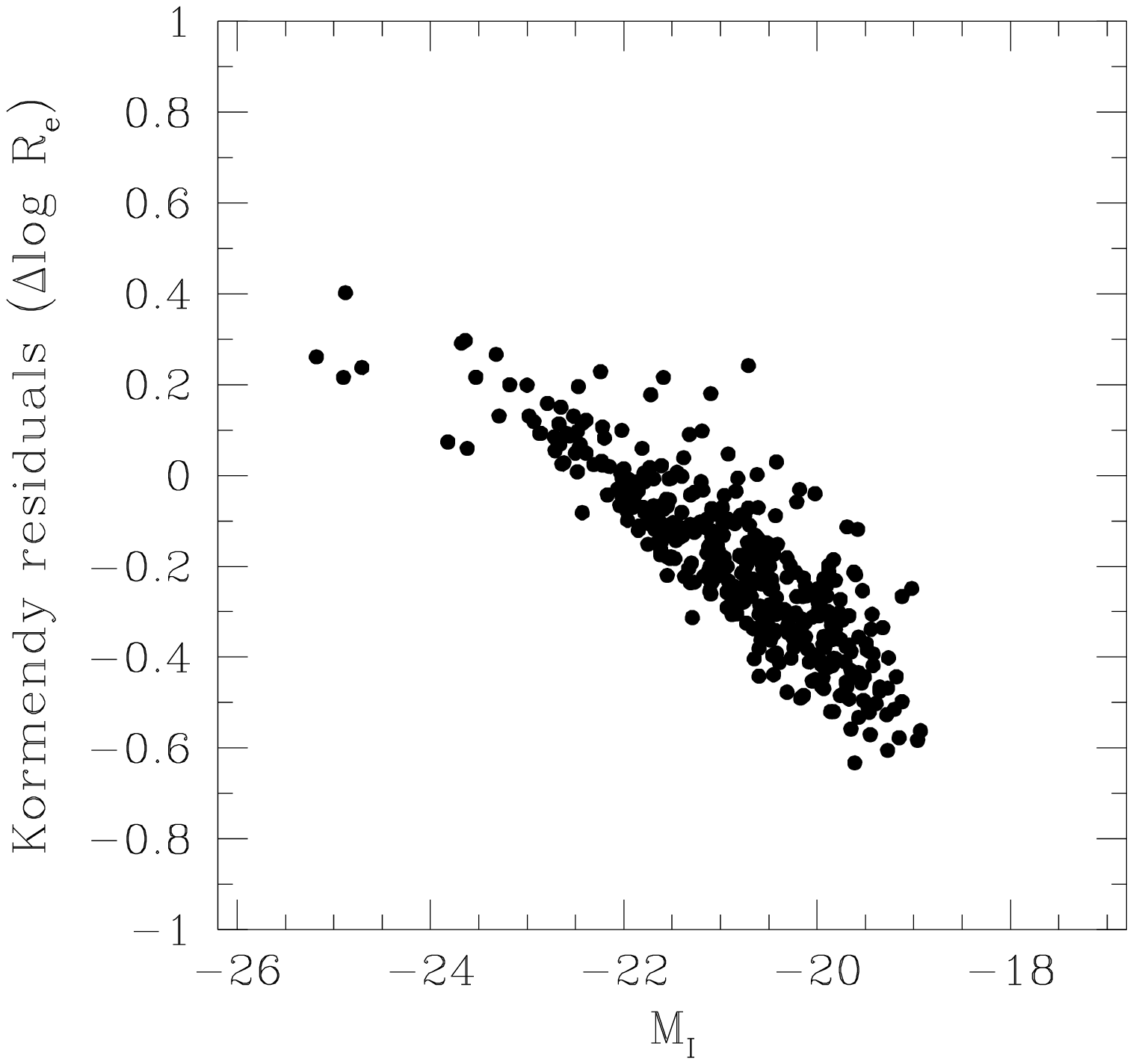}
\figcaption[fig2.ps]{The residuals from the best line fit to the Kormendy
relation (equation 1), plotted as a function of the galaxy magnitude.}

\newpage

\plotone{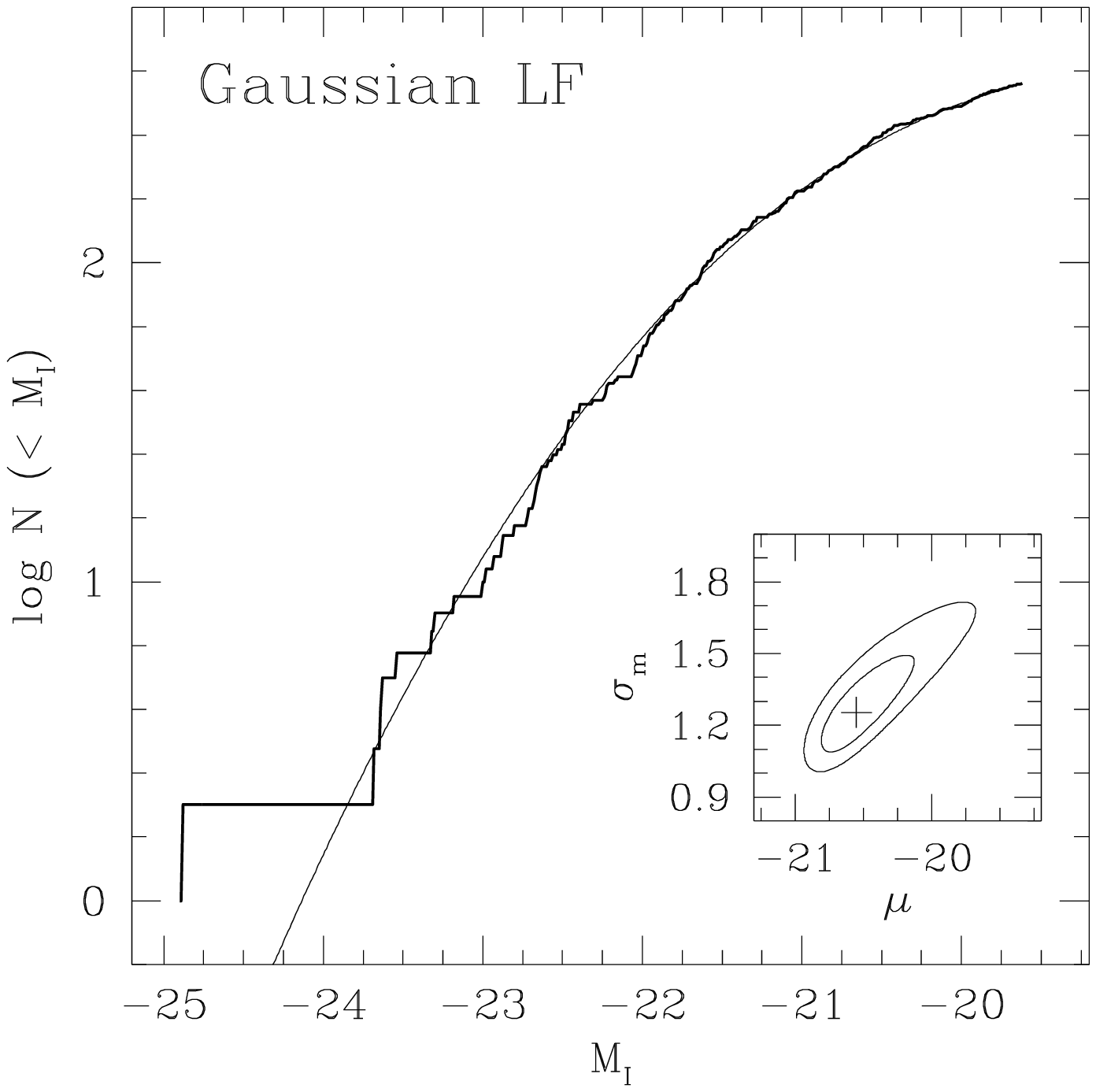}
\figcaption[fig3a.ps]{Maximum likelihood Gaussian luminosity function fit to the
magnitude distribution of the combined sample of early-type galaxies. The 
fitting was performed excluding the brightest galaxy in each cluster. (a) The 
stair-steps line gives the observed integral magnitude distribution, while the 
smooth curve is derived integrating the best fitting Gaussian. The inset show 
the best fitting Gaussian parameters, and the 68\% (1$\sigma$) and 95\% 
(2$\sigma$) joint confidence contours for those parameters. (b) The histogram 
shows the differential magnitude distribution, within bins of width 0.4 mag., 
and the solid line curve shows the best fitting Gaussian. The completeness 
limit for the sample is shown by the vertical dashed line.}

\newpage

\plotone{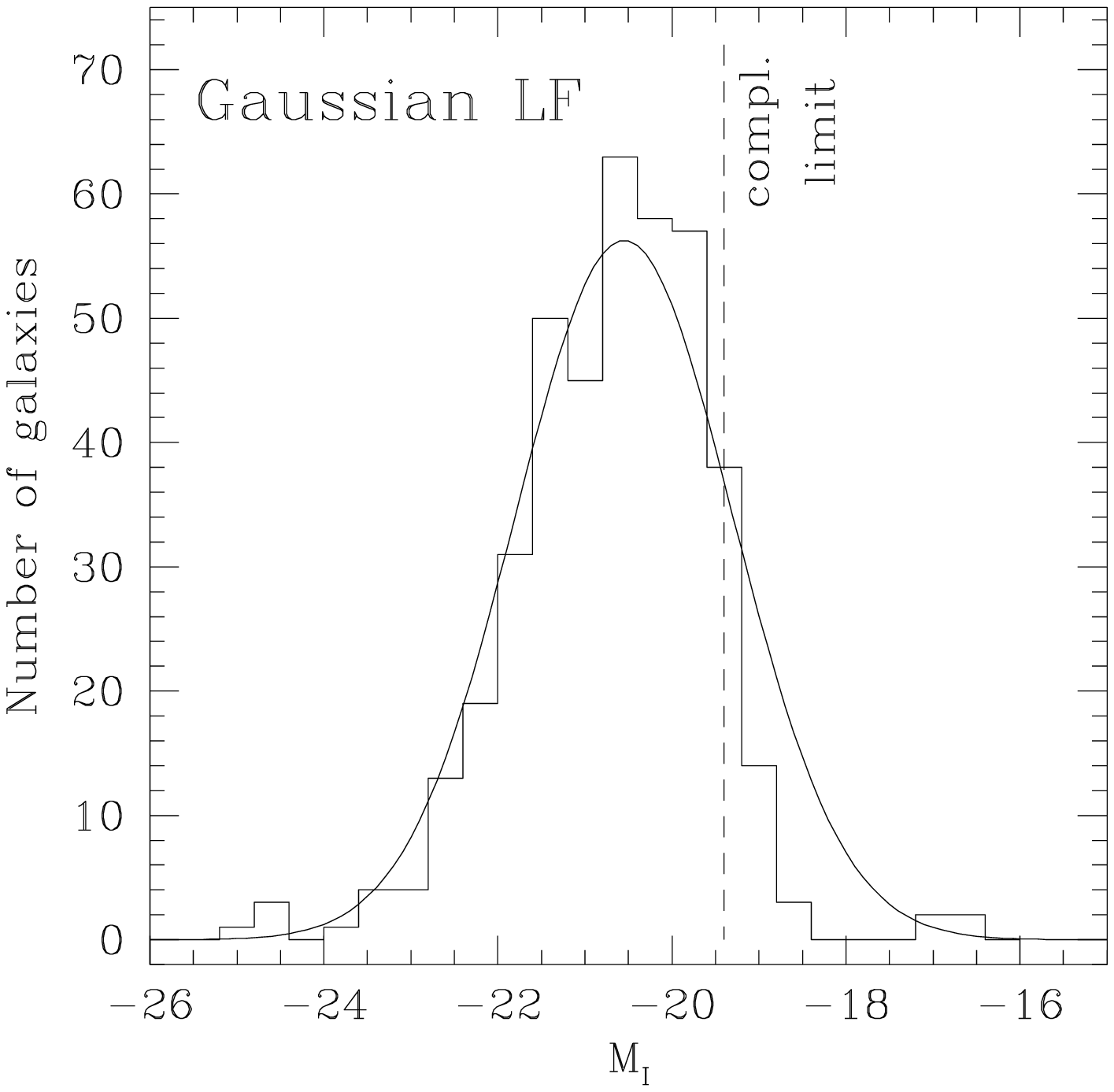}

\newpage

\plotone{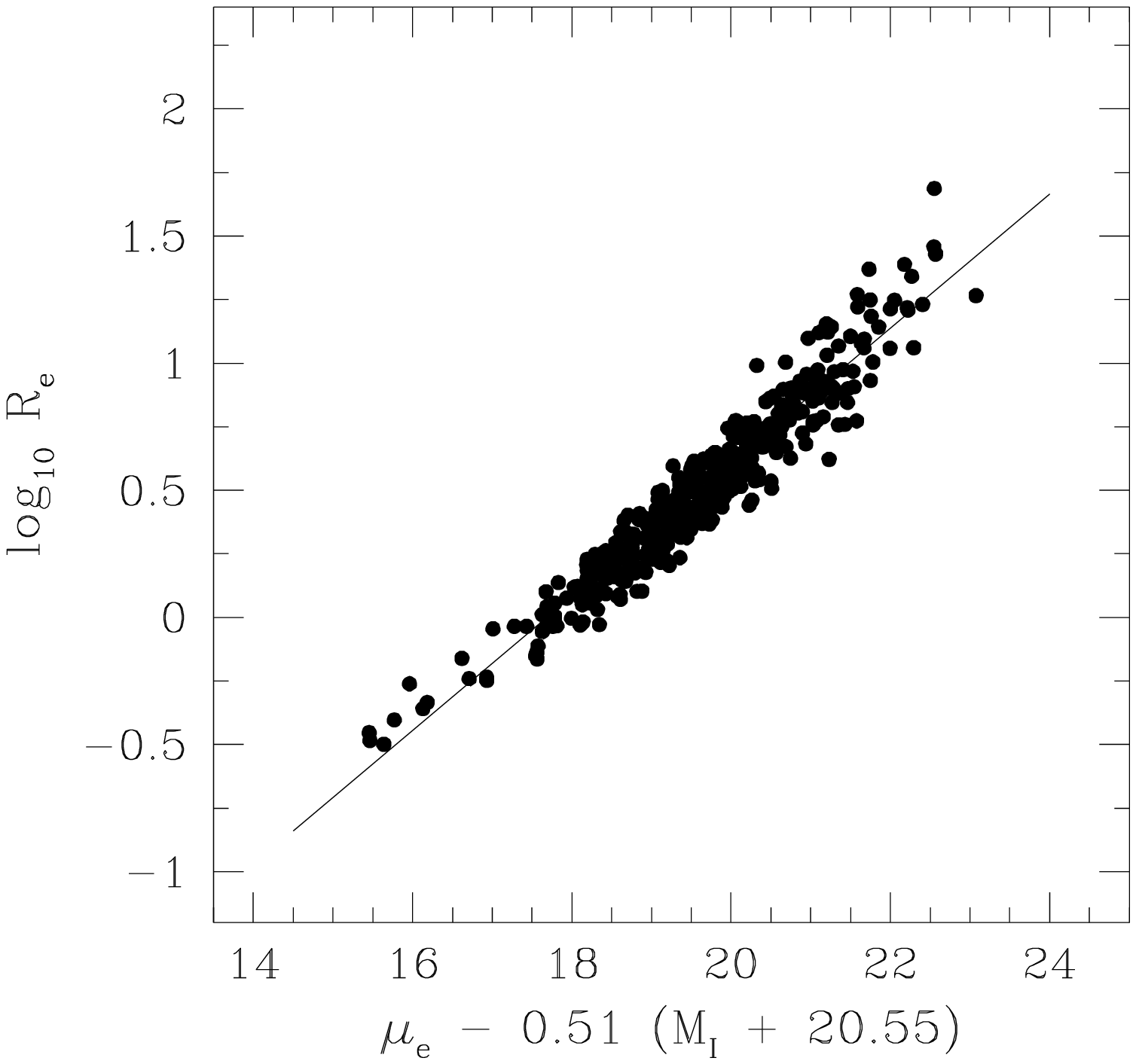}
\figcaption[fig4.ps]{Edge-on view of the PFP.
The same galaxies are plotted as in Fig. 1. The solid line is the 
projection of the best fitting plane (equation 4).}

\newpage

\plotone{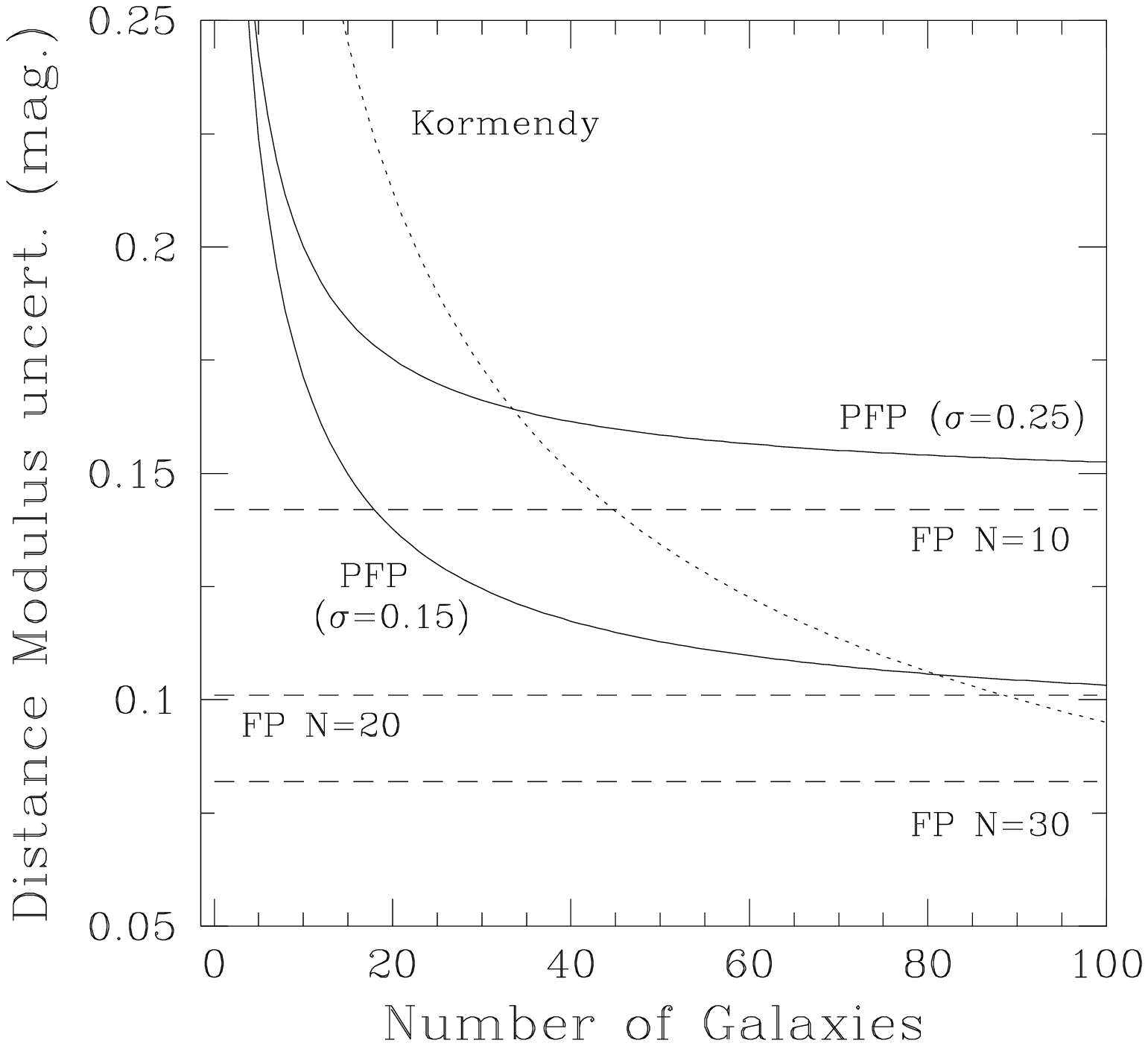}
\figcaption[fig5.ps]{Comparison of the accuracy in a distance modulus estimate, 
as a function of sample size, that can be obtained using the PFP, Kormendy, 
and FP relation. The horizontal dashed lines give the accuracy achieved using 
the FP with a fixed sample size of 10, 20, or 30 galaxies (the scatter in the 
FP is assumed to be 0.085, or 0.43 mag.).
The two solid line curves give the accuracy achieved using the PFP, for two 
different values of the uncertainty associated with the determination of the 
characteristic magnitude, for a scatter in the PFP of 0.096, or 0.48 mag. The 
dotted line gives the accuracy obtained with the Kormendy relation, for a 
scatter of 0.19, or 0.95 mag.}

\end{document}